\documentclass[supp]{new_tlp} 

\newcommand\etc{\emph{etc.}}
\newcommand\bcmdtab{\noindent\bgroup\tabcolsep=0pt%
  \begin{tabular}{@{}p{10pc}@{}p{20pc}@{}}}
\newcommand\ecmdtab{\end{tabular}\egroup}

\usepackage{graphicx, tikz}
\usepackage{amsmath} 

\newcommand{\true}{\mathsf {true}}
\newcommand{\false}{\mathsf {false}}

\newcommand{\B}{{\cal B}}
\newcommand{\C}{{\cal C}}
\newcommand{\D}{{\cal D}}

\newcommand{\lfp}{{\sf lfp}}

\newcommand{\raf}{{\sf raf}}
\newcommand{\qa}{{\sf qa}}
\newcommand{\spl}{{\sf split}}

\newcommand{\PDef}{\mathit{Q}}
\newcommand{\atomconstraints}{\mathsf{atomconstraints}}
\newcommand{\thresholds}{\mathsf{thresholds}}

\def\ll{[\![}
\def\rr{]\!]}

\newenvironment{SProg}
     {\begin{small}\begin{tt}\begin{tabular}[t]{l}}%
     {\end{tabular}\end{tt}\end{small}}
\def\anno#1{{\ooalign{\hfil\raise.07ex\hbox{\small{\rm #1}}\hfil%
        \crcr\mathhexbox20D}}}

  \title[Tools for Constrained Horn Clause Verification]
        {Analysis and Transformation Tools for Constrained Horn Clause Verification\thanks{The research leading to these
  results has received funding from the European Union 7th
  Framework Programme 
  under grant agreement no.
  318337, ENTRA - Whole-Systems Energy Transparency and the Danish Natural Science Research Council grant NUSA: Numerical and Symbolic Abstractions for Software Model Checking. }
        }

\author[John P. Gallagher and Bishoksan Kafle]
         {John P. Gallagher\\
           Roskilde University, Denmark and IMDEA Software Institute, Madrid, Spain \\
          \email{jpg@ruc.dk}
          \and 
          Bishoksan Kafle\\
         Roskilde University, Denmark\\
          \email{kafle@ruc.dk}}

\jdate{March 2003}
\pubyear{2003}
\pagerange{\pageref{firstpage}--\pageref{lastpage}}
\doi{S1471068401001193}

\newtheorem{lemma}{Lemma}[section]

\begin{document}

\label{firstpage}

\maketitle


  \begin{abstract}
Several techniques and tools have been developed for verification of properties expressed as Horn clauses with constraints over a background theory (CHC).  Current CHC verification tools implement intricate algorithms and are often limited to certain subclasses of CHC problems. Our aim in this work is to investigate the use of a combination of off-the-shelf techniques from the literature in analysis and transformation of Constraint Logic Programs (CLPs) to solve challenging CHC verification problems.  We find that many problems can be solved using a combination of tools based on well-known techniques from abstract interpretation, semantics-preserving transformations, program specialisation and query-answer transformations. This gives insights into the design of automatic, more general CHC verification tools based on a library of components.

  \end{abstract}


\section{Introduction}

CHCs provide a suitable intermediate form for expressing the semantics of a variety of programming languages (imperative, functional, concurrent, \etc) and computational models (state machines, transition systems, big- and small-step operational semantics, Petri nets, \etc).  As a result it has been used as a target language for software verification. Recently there is a growing interest in CHC verification from both the logic programming and software verification communities, and several verification techniques and tools have been developed for CHC. 

Pure CLPs are syntactically and semantically the same as CHC. The main difference is that sets of constrained Horn clauses are not necessarily intended for execution, but rather as specifications. From the point of view of verification, we do not distinguish between CHC and pure CLP.  Much research has been carried out on the analysis and transformation of CLP programs, typically for synthesising efficient programs or for inferring run-time properties of programs for the purpose of debugging, compile-time optimisations or program understanding. In this paper we investigate the application of this research to the CHC verification problem.  

In Section \ref{sec:chc} we define the CHC verification problem.  In Section \ref{sec:reltools} we define basic transformation and analysis components drawn from or inspired by the CLP literature. Section \ref{sec:chc-roles} discusses the role of these components in verification, illustrating them on an example problem. In Section \ref{sec:toolcombination} we construct a tool-chain out of these components and test it on a range of CHC verification benchmark  problems. The results reported represent one of the main contributions of this work.  In Section \ref{sec:related} we propose possible extensions of the basic tool-chain and compare them with related work on CHC verification tool architectures.  Finally in Section \ref{sec:concl} we summarise the conclusions from this work.

\section{Background: The CHC Verification Problem}\label{sec:chc}

A CHC is a first order predicate logic formula of the form 
$\forall(\phi \wedge B_1(X_1) \wedge \ldots \wedge B_k(X_k) \rightarrow  H(X))$ ($k \ge 0$),  where $\phi$ is a conjunction of constraints with respect to some background theory, $X_i, X$  are (possibly empty) vectors of distinct variables, $B_1,\ldots,B_k, H$ are predicate symbols, $H(X)$ is the head of the clause and $\phi \wedge B_1(X_1) \wedge \ldots \wedge B_k(X_k)$ is the body.   Sometimes the clause is written $H(X) \leftarrow \phi \wedge B_1(X_1),\ldots, B_k(X_k)$ and in concrete examples it is written in the form \texttt{H :- $\phi$, B$_1$(X$_1$),$\ldots$,B$_k$(X$_k$)}.  In examples,  predicate symbols start with lowercase letters while we use uppercase letters for variables.

We assume here that the constraint theory is linear arithmetic with relation symbols $\le$, $\ge$, $>$, $<$ and $=$ and that there is a distinguished predicate symbol $\false$ which is interpreted as false.  In practice the predicate $\false$ only occurs in the head of clauses; we call clauses whose head is $\false$ \emph{integrity constraints}, following the terminology of deductive databases. Thus the formula $\phi_1 \leftarrow \phi_2 \wedge B_1(X_1),\ldots, B_k(X_k)$ is equivalent to the formula $\false \leftarrow  \neg\phi_1 \wedge \phi_2 \wedge B_1(X_1),\ldots, B_k(X_k)$.  The latter might not be a CHC but can be converted to an equivalent set of CHCs by transforming the formula $\neg\phi_1$ and distributing any disjunctions that arise over the rest of the body.  
For example, the formula 
\texttt{X=Y :-  p(X,Y)}
is equivalent to the set of CHCs \texttt{
false :- X>Y, p(X,Y)} and \texttt{false :- X<Y, p(X,Y)}.
Integrity constraints can be viewed as safety properties.  If a set of CHCs encodes the behaviour of some system, the bodies of integrity constraints represent unsafe states.  Thus proving safety consists of showing that the bodies of integrity constraints are false in all models of the CHC clauses.

\paragraph{The CHC verification problem.} To state this more formally, given a set of CHCs $P$, the CHC verification problem is to check whether there exists a model of $P$. We restate this property in terms of the derivability of the predicate $\false$.

\begin{lemma} $P$ has a model if and only if $P \not\models \false$.
\end{lemma}

\begin{proof}
Let us write $I(F)$ to mean that interpretation $I$ satisfies $F$ ($I$ is a model of $F$). \[
\begin{array}{lll}
P \not\models \false &\equiv & \exists I.(I(P) \textrm{~and~} \neg I(\false))\\
&\equiv&  \exists I. I(P)~~~\textit{(since~} \neg I(\false) \textit{~is~true~by~defn.~of~} \false\textit{)}\\
&\equiv& P \textrm{~has~a~model.}\\
\end{array}\]
\end{proof}
\noindent
Obviously any model of $P$ assigns false to the bodies of integrity constraints.

The verification problem can be formulated deductively rather than model-theoretically. Let the relation $P \vdash A$ denote that $A$ is derivable from $P$ using some proof procedure.  If the proof procedure is sound and complete then $P \not\models A$ if and only if $P \not\vdash A$. So the verification problem is to show (using CLP terminology) that the computation of the goal $\leftarrow \false$ in program $P$ does not succeed using a complete proof procedure.  Although in this work we follow the model-based formulation of the problem, we exploit the equivalence with the deductive formulation, which underlies, for example,  the query-answer transformation and specialisation techniques to be presented.

\subsection{Representation of Interpretations}
An interpretation of a set of CHCs is represented as a set of \emph{constrained facts} of the form $A \leftarrow \mathcal{C}$ where $A$ is an atomic formula $p(Z_1,\ldots,Z_n)$ where $Z_1,\ldots,Z_n$ are distinct variables and $\mathcal{C}$ is a constraint over $Z_1,\ldots,Z_n$. If $\mathcal{C}$ is $\true$ we write $A \leftarrow$ or just $A$. The constrained fact $A \leftarrow \mathcal{C}$ is shorthand for the set of variable-free facts $A\theta$ such that $\mathcal{C}\theta$ holds in the constraint theory, and an interpretation $M$ denotes the set of all facts denoted by its elements; $M$ assigns true to exactly those facts.  
$M_1 \subseteq M_2$ if the set of denoted facts of $M_1$ is contained in the set of denoted facts of $M_2$.  

\paragraph{Minimal models.} A model of a set of CHCs is an interpretation that satisfies each clause.  There exists a minimal model with respect to the subset ordering, denoted $M\ll P \rr$ where $P$ is the set of CHCs. $M\ll P\rr$ can be computed as the least fixed point ($\lfp$) of an immediate consequences operator, $T_P^{\mathcal{C}}$, which is an extension of the standard $T_P$ operator from logic programming, extended to handle constraints \cite{JaffarM94}.  Furthermore $\lfp(T^{\mathcal{C} }_{P} )$ can be computed as the limit of the ascending sequence of interpretations \emph{$\emptyset, T^{\mathcal{C} }_{P}(\emptyset), T^{\mathcal{C} }_{P}(T^{\mathcal{C} }_{P}(\emptyset)), \ldots$}. For more details, see \cite{JaffarM94}. This sequence provides a basis for abstract interpretation of CHC clauses.   

\paragraph{Proof by over-approximation of the minimal model.} It is a standard theorem of CLP that the minimal model $M\ll P \rr$ is equivalent to the set of atomic consequences of $P$.  That is, $P \models p(v_1,\ldots,v_n)$ if and only if $p(v_1,\ldots,v_n) \in M\ll P \rr$. Therefore, the CHC verification problem for $P$ is equivalent to checking that $\false  \not\in M\ll P \rr$. It is sufficient to find a set of constrained facts $M'$ such that $M\ll P \rr \subseteq M'$, where $\false  \not\in M'$.  This technique is called proof by over-approximation of the minimal model.

\section{Relevant tools for CHC Verification}
\label{sec:reltools}
In this section, we give a brief description of some relevant tools borrowed from the literature in analysis and transformation of CLP.  

\paragraph{Unfolding.} Let $P$ be a set of CHCs and $c_0 \in P$ be $H(X) \leftarrow \B_1, p(Y),\B_2$ where $\B_1, \B_2$ are possibly empty conjunctions of atomic formulas and constraints. Let $\{c_1,\ldots,c_m\}$ be the set of clauses of $P$ that have predicate $p$ in the head, that is, $c_i = p(Z_i) \leftarrow \D_i$, where the variables of these clauses are standardised apart from the variables of $c_0$ and from each other.  Then the result of unfolding $c_0$ on $p(Y)$ is the set of CHCs $P' = P \setminus \{c_0\} \cup \{c'_1,\ldots,c'_m\}$ where $c'_i = H(X) \leftarrow \B_1, Y=Z_i,\D_i,\B_2$. The equality $Y=Z_i$ stands for the conjunction of the equality of the respective elements of the vectors $Y$ and $Z_i$.
It is a standard result that unfolding a clause in $P$ preserves $P$'s minimal model \cite{Pettorossi-Proietti}.  In particular, $P \models \false \equiv P' \models \false$.

\paragraph{Specialisation.}  A set of CHCs $P$ can be specialised with respect to a query.  Assume $A$ is an atomic formula;  then we can derive a set $P_A$ such that $P \models A \equiv P_A \models A$.  $P_A$ could be simpler than $P$, for instance, parts of $P$ that are irrelevant to $A$ could be omitted in $P_A$. In particular, the CHC verification problem for $P_{\false}$ and $P$ are equivalent.  There are many techniques in the CLP literature for deriving a specialised program $P_A$.  Partial evaluation is a well-developed method \cite{gallagher:pepm93,Leuschel98a}. 

We make use a form of specialisation know as forward slicing,  more specifically redundant argument filtering \cite{LeuschelS96a}, in which predicate arguments can be removed if they do not affect a computation.  Given a set of CHCs $P$ and a query $A$, denote by $P^{\raf}_A$ the program obtained by applying the RAF algorithm from  \cite{LeuschelS96a} with respect to the goal $A$.  We have the property that $P \models A \equiv P^{\raf}_A \models A$ and in particular that $P \models \false \equiv P^{\raf}_{\false} \models \false$.

\paragraph{Query-answer transformation.} Given a set of CHCs $P$ and an atomic query $A$, the query-answer transformation of $P$ with respect to $A$ is a set of CHCs which simulates the computation of the goal $\leftarrow A$ in $P$, using a left-to-right computation rule. 
Query-answer transformation is a generalisation of the magic set transformations for Datalog. For each predicate $p$,  two new predicates $p_{ans}$ and $p_{query}$ are defined. For an atomic formula $A$, $A_{ans}$ and $A_{query}$ denote the replacement of $A$'s predicate symbol  $p$ by $p_{ans}$ and $p_{query}$ respectively.  Given a program $P$ and query $A$, the idea is to derive a program $P^{\qa}_A$ with the following property  $P \models A$ iff $P^{\qa}_A \models A_{ans}$. The $A_{query}$ predicates represent calls in the computation tree generated during the execution of the goal. For more details see \cite{Debray-Ramakrishnan-94,Gallagher-deWaal-LOPSTR92,Codish-Demoen}. In particular, $P^{\qa}_{\false} \models \false_{ans} \equiv P \models \false$, so we can transform a CHC verification problem to an equivalent CHC verification problem on the query-answer program generated with respect to the goal $\leftarrow \false$.

\paragraph{Predicate splitting.}  Let $P$ be a set of CHCs and let $\{c_1,\ldots,c_m\}$ be the set of clauses in $P$ having some given predicate $p$ in the head, where $c_i = p(X) \leftarrow \D_i$. Let $C_1,\ldots,C_k$ be some partition of $\{c_1,\ldots,c_m\}$, where $C_j = \{c_{j_1},\ldots,c_{j_{n_j}}\}$. Define $k$ new predicates $p_1 \ldots p_k$, where $p_j$ is defined by the bodies of clauses in partition $C_j$, namely $\PDef^j = \{p_j(X) \leftarrow \D_{j_1}, \ldots, p_j(X) \leftarrow \D_{j_{n_j}}\}$.  Finally, define $k$ clauses $C_p = \{p(X) \leftarrow p_1(X), \dots,p(X) \leftarrow p_k(X)\}$.  Then we define a splitting transformation as follows. 
\begin{enumerate}
\item
Let $P' = P\setminus \{c_1,\ldots,c_m\} \cup C_p \cup \PDef^1 \cup \ldots \cup \PDef^k$.
\item
Let $P^{\spl}$ be the result of unfolding every clause in $P'$ whose body contains $p(Y)$ with the clauses $C_p$.

\end{enumerate}
In our applications, we use splitting to create separate predicates for  clauses for a given predicate whose constraints are mutually exclusive. For example, given the clauses \texttt{new3(A,B) :- A=<99, new4(A,B)} and \texttt{new3(A,B) :- A>=100, new5(A,B)}, we produce two new predicates, since the constraints \texttt{A=<99} and \texttt{A>=100} are disjoint. The new predicates are defined by clauses \texttt{new3$_1$(A,B) :- A=<99, new4(A,B)} and \texttt{new3$_2$(A,B) :- A>=100, new5(A,B)}, and all calls to \texttt{new3} throughout the program are unfolded using these new clauses. 
Splitting has been used in the CLP literature to improve the precision of program analyses, for example in \cite{SerebrenikS01a}. In our case it improves the precision of the convex polyhedron analysis discussed below, since separate polyhedra will be maintained for each of the disjoint cases.  The correctness of splitting can be shown using standard transformations that preserve the minimal model of the program (with respect to the predicates of the original program) \cite{Pettorossi-Proietti}.  Assuming that the predicate $\false$ is not split, we have that $P \models \false \equiv P^{\spl} \models \false$.

\paragraph{Convex polyhedron approximation.} 
Convex polyhedron analysis \cite{Cousot-Halbwachs-78} is a program analysis technique based on abstract interpretation \cite{DBLP:conf/popl/CousotC77}. When applied to a set of CHCs $P$ it constructs an over-approximation $M'$ of the minimal model of $P$, where $M'$ contains at most one constrained fact $p(X) \leftarrow \mathcal{C}$ for each predicate $p$. The constraint $\mathcal{C}$ is a conjunction of linear inequalities, representing a convex polyhedron. 
The first application of convex polyhedron analysis to CLP was by  \citeN{Benoy-King-LOPSTR96}.
Since the domain of convex polyhedra contains infinite increasing chains, the use of a widening operator is needed to ensure convergence of the abstract interpretation. Furthermore much research has been done on improving the precision of widening operators.  One technique is known as widening-upto, or widening with thresholds \cite{Halbwachs-94}.  

Recently, a technique for deriving more effective thresholds was developed \cite{Lakhdar-ChaouchJG11}, which we have adapted and found to be effective in experimental studies. The thresholds are computed by the following method.  Let $T^{\C}_P$ be the standard immediate consequence operator for CHCs, that is, $T^{\C}_P(I)$ is the set of constrained facts that can be derived in one step from  a set of constrained facts $I$.  Given a constrained fact $p(\bar Z) \leftarrow \C$, define $\atomconstraints(p(\bar Z) \leftarrow \C)$ to be the set of constrained facts $\{p(\bar Z) \leftarrow C_i \mid \C = C_1 \wedge \ldots \wedge C_k, 1 \le i \le k)\}$.  The function $\atomconstraints$ is extended to interpretations by $\atomconstraints(I) = \bigcup_{p(\bar Z) \leftarrow \C \in I}\{\atomconstraints(p(\bar Z) \leftarrow \C)\}$.

Let $I_{\top}$ be the interpretation consisting of the set of constrained facts $p(\bar{Z}) \leftarrow \true$ for each predicate $p$. We perform three iterations of $T^{{\C}}_P$ starting with $I_{\top}$ (the first three elements of a ``top-down" Kleene sequence) and then extract the atomic constraints. That is, $\thresholds$ is defined as follows.
\[
\thresholds(P) = \atomconstraints(T^{{\C}(3)}_P(I_{\top}))
\]
\noindent
A difference from the method in \cite{Lakhdar-ChaouchJG11} is that we use the concrete semantic function $T^{\C}_P$ rather than the abstract semantic function when computing thresholds.  The set of threshold constraints represents an attempt to find useful predicate properties and when widening they help to preserve invariants that might otherwise be lost during widening.  See \cite{Lakhdar-ChaouchJG11} for further details.  Threshold constraints that are not invariants are simply discarded during widening.

\section{The role of CLP tools in verification}\label{sec:chc-roles}

The techniques discussed in the previous section play various roles.
The convex polyhedron analysis, together with the helper tool to derive threshold constraints, constructs an approximation of the minimal model of a CHC theory.  If $\false$ (or $\false_{ans}$) is not in the approximate model, then the verification problem is solved.  Otherwise the problem is not solved;  in effect a ``don't know" answer is returned. We have found that polyhedron analysis alone is seldom precise enough to solve non-trivial CHC verification problems; in combination with the other tools, it is very effective.

Unfolding can improve the structure of a program, removing some cases of mutual recursion, or propagating constraints upwards towards the integrity constraints, and can improve the precision and performance of convex polyhedron analysis.

Specialisation can remove parts of theories not relevant to the verification problem, and can also propagate constraint downwards from the integrity constraints. Both of these have a beneficial effect on performance and precision of polyhedron analysis.

Analysis of a query-answer program (with respect to $\false$) is in effect the search for a derivation tree for $\false$.   Its effectiveness in CHC verification problems is variable. It can sometimes worsen performance since the query-answer transformed program is larger and contains more recursive dependencies than the original. On the other hand, one seldom loses precision and it is often more effective in allowing constraints to be propagated upwards (through the $ans$ predicates) and downwards (through the $query$ predicates). 

\begin{figure}
\centering
\begin{verbatim}
new6(A,B) :- B=<99.                    new4(A,B) :- C=1+A,D=1+B,A>=50,new3(C,D).
new5(A,B) :- B>=101.                   new3(A,B) :- A=<99, new4(A,B).
new5(A,B) :- B=<100, new6(A,B).        new3(A,B) :- A>=100, new5(A,B).
new4(A,B) :- C=1+A, A=<49, new3(C,B).  false :- A=0, B=50, new3(A,B).
\end{verbatim}

\caption{The example program 
\texttt{MAP-disj.c.map.pl} }
\label{prg:proi1}
\end{figure}

\subsection{Application of the tools}

We illustrate the tools on a running example (Figure \ref{prg:proi1}), one of the benchmark suite of the VeriMAP system \citeN{DBLP:conf/cilc/AngelisFPP13}. The result of applying unfolding is shown in Figure \ref{prg:proi1_unfolded} (omitting the definitions of the unfolded predicates \texttt{new4, new5} and \texttt{new6}, which are no longer reachable from \texttt{false}).  The unfolding strategy we adopt is the following:  the predicate dependency graph of a program consists of the set of edges $(p,q)$ such that there is clause where $p$ is the predicate of the head and $q$ is a predicate occurring in the body.  We perform a depth-first search of the predicate dependency graph, starting from \texttt{false}, and identify the backward edges, namely those edges $(p,q)$ where $q$ is an ancestor of $p$ in the depth-first search. We then unfold every body call whose predicate is not at the end of a backward edge. In Figure \ref{prg:proi1}, we thus unfold calls to \texttt{new4, new5} and \texttt{new6}.

\begin{figure}
\centering
\begin{tabular}{ll}
\begin{SProg}
false :- A=0,  B=50, new3(A,B).\\
new3(A,B) :- A=<99, C = 1+A, A=<49, new3(C,B).\\
new3(A,B) :- A=<99, C = 1+A, D = 1+B, A>=50, new3(C,D).\\
new3(A,B) :- A>=100, B>=101.\\
new3(A,B) :-  A>=100,  B=<100, B=<99.\\
\end{SProg}
\end{tabular}
\caption{Result of unfolding \texttt{MAP-disj.c.map.pl} }
\label{prg:proi1_unfolded}
\end{figure}

\begin{figure}
\centering
\begin{tabular}{ll}
\begin{SProg}
false\_ans :-
      false\_query,
      A=0,
      B=50,
      new3\_ans(A,B).\\
new3\_ans(A,B) :-
      new3\_query(A,B),
      A=<99,
      C = 1+A,
      A=<49,
      new3\_ans(C,B).\\
new3\_ans(A,B) :-
new3\_query(A,B),A=<99,C is 1+A,D is 1+B,
      A>=50,
      new3\_ans(C,D).\\
new3\_ans(A,B) :-
      new3\_query(A,B),
      A>=100,
      B>=101.\\
new3\_ans(A,B) :-
      new3\_query(A,B),
      A>=100,
      B=<100,
      B=<99.\\
new3\_query(A,B) :-
      false\_query,
      A=0,
      B=50.\\
new3\_query(A,B) :-
      new3\_query(C,B),
      C=<99,
      A = 1+C,
      C=<49.\\
new3\_query(A,B) :-
      new3\_query(C,D),
      C=<99,
      A = 1+C,
      B = 1+D,
      C>=50.\\
false\_query.
\end{SProg}
\end{tabular}
\caption{The query-answer transformed program for
program of Figure \ref{prg:proi1_unfolded}}
\label{prg:proi1-qa}
\end{figure}
The query-answer transformation is applied to the program in Figure \ref{prg:proi1_unfolded}, with respect to the goal \texttt{false} resulting in the program shown in Figure \ref{prg:proi1-qa}.  The model of the predicate \texttt{new3\_query} corresponds to those calls to \texttt{new3} that are reachable from the call in the integrity constraint.  Explicit representation of the query predicates permits more effective propagation of constraints from the integrity clauses during model approximation.

The splitting transformation is now applied to the program in Figure \ref{prg:proi1-qa}.  We do not show the complete program, which contains 30 clauses.  Figure \ref{prg:proi1-qa-split} shows the split definition of  \texttt{new3\_query}, which is split since the last two clauses for \texttt{new3\_query} in Figure \ref{prg:proi1-qa} have mutually disjoint constraints, when projected onto the head variables.  
\begin{figure}
\centering
\begin{tabular}{ll}
\begin{SProg}
new3\_query\_\_\_1(A,B) :-
   false\_query\_\_\_1,
   A=0,
   B=50.\\
new3\_query\_\_\_1(A,B) :-
   new3\_query\_\_\_1(C,B),
   C=<99,
   A = 1+C,
   C=<49.\\
new3\_query\_\_\_1(A,B) :-
   new3\_query\_\_\_2(C,B),
   C=<99,
   A = 1+C,
   C=<49.\\

new3\_query\_\_\_2(A,B) :-
   new3\_query\_\_\_1(C,D),
   C=<99,
   A = 1+C,
   B = 1+D,
   C>=50.\\
new3\_query\_\_\_2(A,B) :-
   new3\_query\_\_\_2(C,D),
   C=<99,
   A = 1+C,
   B = 1+D,
   C>=50.\\
\end{SProg}
\end{tabular}
\caption{Part of the split program for
the program in Figure \ref{prg:proi1-qa}}
\label{prg:proi1-qa-split}
\end{figure}

A convex polyhedron approximation is then computed for the split program, after computing threshold constraints for the predicates.  The resulting approximate model is shown in Figure \ref{cha-out} as a set of constrained facts.  Since the model does not contain any constrained fact for \texttt{false\_ans} we conclude that \texttt{false\_ans} is not a consequence of the split program.  Hence, applying the various correctness results for the unfolding, query-answer and splitting transformations, \texttt{false} is not a consequence of the original program.
\begin{figure}
\centering
\begin{tabular}{ll}
\begin{SProg}
false\_query\_\_\_1 :- []\\
new3\_query\_\_\_1(A,B) :- [1*A>=0,-1*A>= -50,1*B=50]\\
new3\_query\_\_\_2(A,B) :- [1*A>=51,-1*A>= -100,1*A+ -1*B=0]\\
\end{SProg}
\end{tabular}
\caption{The convex polyhedral approximate model for the split program}
\label{cha-out}
\end{figure}

\begin{figure}[h!]
\begin{center}
\begin{tikzpicture}
\draw[thick,  color=blue] (0.5,0) rectangle (12.8,4); 
\begin{scope}
 \node at (8,3.8) {\it RAF -- Redundant Argument Filtering};
  \node at (8,3.4) {\it FU -- Forward Unfolding };
   \node at (8,3.0) {\it QA -- Query Answer Transformation };
   
   \node at (4,1.0) {\it PS -- Predicate Splitting };
   \node at (4,0.6) {\it TC -- Threshold Constraint };
   \node at (4,0.2) {\it CHA -- Convex Hull Analyzer };
    

 \node at (2,3.8) {CHC Program P};
 \draw[->] (1.5,3.5) -- (1.5,2.5);

\draw[blue] (.8,1.5) rectangle (2.0,2.5); 
\node at (1.5,2) {RAF}; 

 \node at (3.3,2.0) {FU};
 \draw[->] (2.0,2) -- (2.7,2);
\draw[blue] (2.7,1.5) rectangle (3.9,2.5); 

\draw[blue] (4.6,1.5) rectangle (5.8,2.5); 
\node at (5.2,2) {QA}; 
 \draw[->] (3.9,2) -- (4.6,2);
 
 \draw[blue] (6.5,1.5) rectangle (7.7,2.5); 
\node at (7.1,2) {PS}; 
 \draw[->] (5.8,2) -- (6.5,2);
 
  \draw[blue] (8.5,1.5) rectangle (9.7,2.5); 
\node at (9.1,2) {TC}; 
 \draw[->] (7.7,2) -- (8.5,2);

 \draw[->] (9.7,2) -- (10.4,2);

 \node at (12,2.3) {Safe};
 \draw[->] (11.6,2) -- (12.5,2);

    \node at (11.8,1) {unknown };
    \draw[->] (11,1.5) -- (11,0.5);

\draw[blue] (10.4,1.5) rectangle (11.6,2.5); 
\node at (11,2) {CHA }; 
\end{scope}
\end{tikzpicture}
\end{center}
\caption{\it The basic tool chain for CHC verification.}
\label{fig:toolchain}
\end{figure}
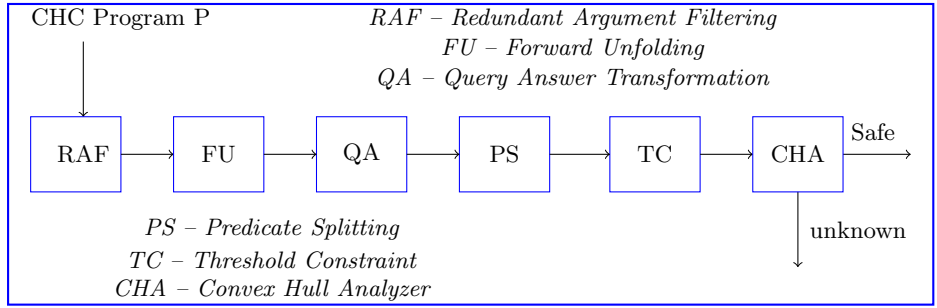

\paragraph{Discussion of the example.} Application of the convex polyhedron tool to the original, or the intermediate programs, does not solve the problem; all the transformations are needed in this case, apart from redundant argument filtering, which only affects efficiency. The ordering of the tool-chain can be varied somewhat, for instance switching query-answer transformation with splitting or unfolding.  In our experiments we found the ordering in Figure \ref{fig:toolchain}
to be the most effective.

The model of the query-answer program is finite for this example.  However, the problem is essentially the same if the constants are scaled;  for instance we could replace 50 by 5000, 49 by 4999, 100 by 10000 and 101 by 10001, and the problem is essentially unchanged.  We noted that some CHC verification tools applied to this example solve the problem, but essentially by enumeration of the finite set of values encountered in the search.  Such a solution does not scale well. On the other hand the polyhedral abstraction shown above is not an enumeration; an essentially similar polyhedron abstraction is generated for the scaled version of the example, in the same time.  The VeriMAP tool \cite{DBLP:conf/cilc/AngelisFPP13} also handles the original and scaled versions of the example in the same time.

\section{Combining off-the-shelf tools: Experiments}
\label{sec:toolcombination}

The motivation for our tool-chain, summarised in Figure \ref{fig:toolchain}, comes from our example program, which is a simple yet challenging program.  We applied the tool-chain to a number of benchmarks from the literature, taken mainly from the repository of Horn clause benchmarks in SMT-LIB2 (https://svn.sosy-lab.org/software/sv-benchmarks/trunk/clauses/) and other sources including \cite{DBLP:journals/tplp/GangeNSSS13} and some of the VeriMap benchmarks \cite{DBLP:conf/cilc/AngelisFPP13}. We selected these examples because many of them are considered challenging because they cannot be solved by one or more of the state-of-the-art-verification tools discussed below. Programs taken from the SMT-LIB2 repository are first translated to CHC form. 
The results are summarised in Table \ref{tbl:experiments}.

In Table \ref{tbl:experiments}, columns Program and Result  respectively represent the benchmark program and the results of verification using our tool combination.  Problems marked with (*) could not be handled by our tool-chain since they contain numbers which do not fit in 32 bits, the limit of our Ciao Prolog implementation.
whereas problems marked with (**) are solvable by simple ad hoc modification of the tool-chain, which we are currently investigating (see Section \ref{sec:concl}).
Problems such as systemc-token-ring.01-safeil.c contain complicated loop structure with large strongly connected components in the predicate dependency graph and our convex polyhedron analysis tool is unable to derive the required invariant.
 However overall  results show that our simple tool-chain begins to compete with advanced tools like HSF \cite{DBLP:conf/tacas/GrebenshchikovGLPR12}, VeriMAP \cite{DBLP:conf/cilc/AngelisFPP13}, TRACER \cite{DBLP:conf/cav/JaffarMNS12}, \etc  $~$We do not report timings, though all these results are obtained in a matter of seconds, since our tool-chain is not at all optimised, relying on file input-output and the individual components are often prototypes.

\begin{table}[h!]
 \caption {Experiments results on CHC benchmark program}
    \label{tbl:experiments}
\begin{minipage}{\textwidth}
    \begin{tabular}{|l|l|l||l|l|l|}
    \hline
    SN     & Program               & Result                  &SN              & Program                            & Result   \\ \hline
    1        & MAP-disj.c.map.pl            & verified                &17                  &  MAP-forward.c.map.pl          & verified \\ 
    2        & MAP-disj.c.map-scaled.pl           & verified                &18               & tridag.smt2                       & verified \\ 
    3        &t1.pl                        & verified                &19                &  qrdcmp.smt2                   & verified \\ 
    4        & t1-a.pl                   & verified                &20                &  choldc.smt2                     & verified \\ 
    5        & t2.pl                      & verified                 &21                &  lop.smt2                           & verified \\ 
    6        & t3.pl                      & verified           &22                &  pzextr.smt2                      & verified \\ 
    7       & t4.pl                       & verified                 &23                & qrsolv.smt2                       & verified \\ 
   8       & t5.pl                        & verified                 & 24               &   INVGEN-apache-escape-absolute                     &verified                                   \\ 
   9       &  pldi12.pl         & verified        & 25               &    TRACER-testabs15                 &verified                                   \\ 
   10       &  INVGEN-id-build       & verified                      & 26**               & amebsa.smt2                     &verified                                   \\ 
   
    11       &  INVGEN-nested5        & verified        & 27**               &DAGGER-barbr.map.c                     &verified                                   \\ 
     12       &  INVGEN-nested6       & verified        & 28*               & sshsimpl-s3-srvr-1a-safeil.c                     &NOT                                   \\ 
      13       &  INVGEN-nested8       & verified        & 29              &   sshsimpl-s3-srvr-1b-safeil.c                 &NOT                                   \\ 
       14       &  INVGEN-svd-some-loop        & verified        & 30*               &   bandec.smt2                 &NOT                                   \\ 
        15       &  INVGEN-svd1       & verified        & 31               &  systemc-token-ring.01-safeil.c                    &NOT                                   \\ 
                16       &  INVGEN-svd4      & verified        & 32*               & crank.smt2                 &NOT                                   \\  \hline
    \end{tabular}
   \end{minipage}
\end{table}

\begin{figure}[h!]
\begin{center}
\begin{tikzpicture}
\draw[thick,  color=blue] (0.5,0) rectangle (12.8,4); 
\begin{scope}
 \node at (8,3.8) {\it PA -- Predicate Abstraction};
 \node at (2,3.8) {CHC Program P};
 \draw[->] (1.5,3.7) -- (1.5,3.2);

\draw[blue] (.8,3.2) rectangle (2.0,2.8); 
\node at (1.5,3) {RAF}; 
 \draw[->] (1.5,2.8) -- (1.5,2.5);
 
\draw[blue] (.8,1.5) rectangle (2.0,2.5); 
\node at (1.5,2) {FU}; 

 \node at (3.3,2.0) {QA};
 \draw[->] (2.0,2) -- (2.7,2);
\draw[blue] (2.7,1.5) rectangle (3.9,2.5); 

\draw[blue] (4.6,1.5) rectangle (5.8,2.5); 
\node at (5.2,2) {PS}; 
 \draw[->] (3.9,2) -- (4.6,2);
 
 \draw[blue] (6.5,1.5) rectangle (7.7,2.5); 
\node at (7.1,2) {TC}; 
 \draw[->] (5.8,2) -- (6.5,2);
 
  \draw[blue] (8.5,1.5) rectangle (9.7,2.5); 
\node at (9.1,2) {CHA}; 
 \draw[->] (7.7,2) -- (8.5,2);

 \draw[->] (9.7,2) -- (10.4,2);

 \node at (12,2.3) {Safe};
\node at (12,1.7) { CEx.};
 \draw[->] (11.6,2) -- (12.5,2);

 \node at (9.5,0.3) {props };
 \draw[dashed] (11,1.5) -- (11,0.5);
  \draw[dashed] (11,0.5) -- (7.1,0.5);
   \draw[->] (7.1,0.5) -- (7.1,1.5);
   
    \node at (12,1) {unknown };
    \draw[->] (11.2,1.5) -- (11.2,0.5);
    
 \draw[dashed] (10.5,2.5) -- (10.5,3);
  \draw[dashed] (10.5,3) -- (11.3,3);
   \draw[->] (11.3,3) -- (11.3,2.5);

\draw[blue] (10.4,1.5) rectangle (11.6,2.5); 
\node at (11,2) {PA }; 
\end{scope}
\end{tikzpicture}
\end{center}
\caption{\it Future extension of our tool-chain.}
\label{fig:toolchain2}
\end{figure}
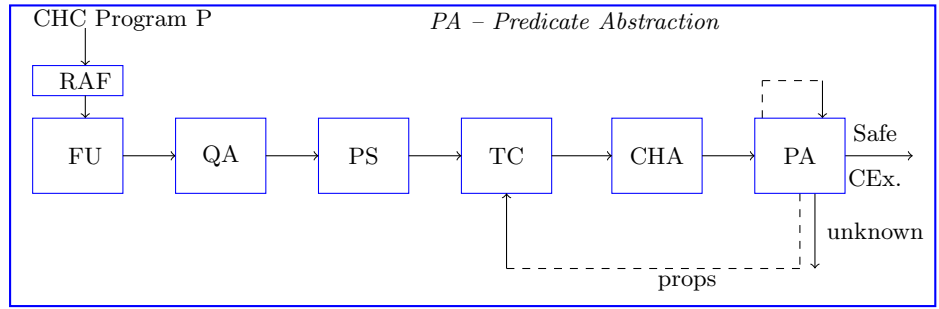

\section{Discussion and Related Work}
\label{sec:related}
The most similar work to ours is by \citeN{DBLP:conf/cilc/AngelisFPP13} which is also based on CLP program transformation and specialisation. They construct a sequence of transformations of $P$, say, $P, P_1, P_2, \ldots,P_k$;  if $P_k$ contains no clause with head $\false$ then the verification problem is solved. A proof of unsafety is obtained if $P_k$ contains a clause $\false \leftarrow$. 
Both our approach and theirs repeatedly apply specialisations preserving the property to be proved. However the difference is that their specialisation techniques are based on unfold-fold transformations, with a sophisticated control procedure  controlling unfolding and generalisation. Our specialisations are restricted to redundant argument filtering and the query-answer transformation, which specialises predicate answers with respect to a goal.  Their test for success or failure is a simple syntactic check, whereas ours is based on an abstract interpretation to derive an over-approximation.  Informally one can say that the hard work in their approach is performed by the specialisation procedure, whereas the hard work in our approach is done by the abstract interpretation.
We believe that our tool-chain-based approach gives more insight into the role of each transformation. 

Work by  \citeN{DBLP:journals/tplp/GangeNSSS13} is a top-town evaluation of CLP programs which records certain derivations and learns only from failed derivations. This helps to prune further derivations and helps to achieve termination in the presence of infinite executions. 
Duality (http://research.microsoft.com/en-us/projects/duality/) and HSF(C) \cite{DBLP:conf/tacas/GrebenshchikovGLPR12} are examples of the CEGAR approach (Counter-Example-Guided Abstraction Refinement). 
This approach can be viewed as property-based abstract interpretation based on a set of properties that is refined on each iteration. The refinement of the properties is the key problem in CEGAR; an abstract proof of unsafety is used to generate properties (often using interpolation) that prevent that proof from arising again. Thus,  abstract counter-examples are successively eliminated.
The relatively good performance of our tool-chain, without any refinement step at all, suggests that finding the right invariants is aided by a tool such as the convex polyhedron solver and the pre-processing steps we applied. 
 In Figure \ref{fig:toolchain2} we sketch possible extensions of our basic tool-chain, incorporating a refinement loop and property-based abstraction.

It should be noted that the query-answer transformation, predicate splitting and unfolding may all cause an blow-up in the program size.  The convex polyhedron analysis becomes more effective as a result, but for scalability we need more sophisticated heuristics controlling these transformations, especially unfolding and splitting, as well as lazy or implicit generation of transformed programs, using techniques such as a fixpoint engine that simulates query-answer programs \cite{DBLP:journals/jlp/Codish99}.

\section{Concluding remarks and future work}
\label{sec:concl}
We have shown that a combination of off-the-shelf tools from CLP transformation and analysis, combined in a sensible way, is surprisingly effective in CHC verification. The component-based approach allowed us to experiment with the tool-chain until we found an effective combination.  This experimentation is continuing and we are confident of making improvements by incorporating other standard techniques and by finding better heuristics for applying the tools. Further we would like to investigate the choice of chain suitable for each  example since more complicated problems can be handled just by altering the chain.  We also suspect from initial experiments that an advanced partial evaluator such as ECCE \cite{DBLP:conf/pepm/LeuschelEVCF06} will play a useful role.
Our results give insights for further development of automatic CHC verification tools.
We would like to combine our program transformation techniques  with abstraction refinement techniques and experiment with the combination. \ \\

\bibliographystyle{acmtrans}
\bibliography{refs}

\end{document}